\documentclass{article}

\begin{document} 

\begin{titlepage}

\begin{center}
(1,9)-Spacetime $\longrightarrow$ (1,3)-Spacetime: Reduction $\Longrightarrow U(1)\times SU(2)\times SU(3)$ \\
\vspace{.25in}
Geoffrey Dixon \\
gdixon@fas.harvard.edu (until July 1999) \\
gdixon@7stones.com (thereafter) \\
\vspace{.25in}
31 January 1999

\vspace{.25in}
A derivation of the standard symmetry  and leptoquark family structure 
is presented that is more straightforward than 
a previous derivation \textbf{[1]}.
\end{center}

\end{titlepage}

Facts and Notation: 
\begin{itemize}

\item \textbf{O} - octonions: nonassociative, noncommutative, basis $\{1=e_{0}, e_{1}, ..., e_{7}\}$;

\item \textbf{Q} - quaternions: associative, noncommutative, basis $\{1=q_{0}, q_{1}, q_{2}, q_{3}\}$; 
\item \textbf{C} - complex numbers: associative, commutative, basis $\{1, i\}$; 

\item \textbf{R} - real numbers. 

\item $\textbf{K}_{L}, \textbf{K}_{R}$ - the adjoint algebras of left and right actions of an 
algebra \textbf{K} on itself.

\item \textbf{K}(2) - 2x2 matrices over the algebra \textbf{K} (to be identified with Clifford algebras);

\item $\mathcal{CL}(p,q)$ - the Clifford algebra of the real spacetime with signature (p+,q-);

\item $^{2}\textbf{K}$ - 2x1 matrices over the algebra \textbf{K} (to be identified with spinor spaces);

\item $\textbf{O}_{L}$ and $\textbf{O}_{R}$ are identical, 
isomorphic to \textbf{R}(8) (8x8 real matrices), \\
64-dimensional bases are of the form $1, \, e_{La}, \, e_{Lab}, \, e_{Labc}$, or 
$1, \, e_{Ra}, \, e_{Rab}, \, e_{Rabc}$, where, for example, if $x \in \textbf{O}$, then 
$e_{Lab}[x] \equiv e_{a}(e_{b}x)$, and $e_{Rab}[x] \equiv (xe_{a})e_{b}$ (see \textbf{[1]});

\item $\textbf{Q}_{L}$ and $\textbf{Q}_{R}$ are distinct, 
both isomorphic to \textbf{Q}, bases \\
$\{1=q_{L0}, q_{L1}, q_{L2}, q_{L3}\}$ and $\{1=q_{R0}, q_{R1}, q_{R2}, q_{R3}\}$;

\item $\textbf{C}_{L}$ and $\textbf{C}_{R}$ are identical, 
both isomorphic to \textbf{C} (so we only need use \textbf{C} itself);

\item \textbf{P} = $\textbf{C}\otimes\textbf{Q}$, 8-dimensional;

\item $\textbf{P}_{L}$ = $\textbf{C}_{L}\otimes\textbf{Q}_{L}$, 
isomorphic to $\textbf{C}(2) \simeq \mathcal{CL}(3,0) \simeq \textbf{C} \otimes \mathcal{CL}(0,2)$ 
(\textbf{P} is the spinor space of $\textbf{P}_{L}$, consisting of a pair of Pauli spinors; 
the doubling is due to the internal action of $\textbf{Q}_{R}$, which commutes with 
$\textbf{P}_{L}$ actions);

\item \textbf{T} = $\textbf{C}\otimes\textbf{Q}\otimes\textbf{O}$, 64-dimensional;

\item $\textbf{T}_{L}$ = $\textbf{C}_{L}\otimes\textbf{Q}_{L}\otimes\textbf{O}_{L}$, 
isomorphic to $\textbf{C}(16) \simeq \mathcal{CL}(0,9) \simeq \textbf{C} \otimes \mathcal{CL}(0,8)$ 
(as was the case with \textbf{P}, the algebra \textbf{T} is the spinor space of $\textbf{T}_{L}$, 
its dimension twice what is expected due to the internal action of $\textbf{Q}_{R}$, the only 
part of $\textbf{T}_{R}$ missing from $\textbf{T}_{L}$);

\item $\textbf{P}_{L}(2) \simeq \textbf{C}(4) \simeq 
\textbf{C}\otimes\mathcal{CL}(1,3)$, the Dirac algebra of (1,3)-spacetime 
(the major difference being that the spinor space, $^{2}\textbf{P}$, 
contains an extra internal $SU(2)$ degree of freedom associated with $\textbf{Q}_{R}$);

\item $\textbf{T}_{L}(2) \simeq \textbf{C}(32) \simeq \textbf{C}\otimes\mathcal{CL}(1,9)$, 
the Dirac algebra of (1,9)-spacetime (spinor space $^{2}\textbf{T}$; one internal $SU(2)$).

\end{itemize}
% -----------------------------------------------------------------------------------------
\newpage

Some Lie algebras and their bases:

\begin{itemize}

\item $so(7)$ - \{$e_{Lab}$: a,b = 1,...,7\};

\item $so(6)$ - \{$e_{Lpq}$: p,q = 1,...,6\};

\item $LG_{2}$ - \{$e_{Lab} - e_{Lcd}$: $e_{a}e_{b} - e_{c}e_{d} = 0$, a,b,c,d = 1,...,7\};

\item $LG_{2}$ explicitly ($LG_{2}$ is the 14-d Lie algebra of $G_{2}$, the automorphism 
group of \textbf{O}): 
\begin{center}
\( \begin{array}{ll}
e_{L24} - e_{L56}, & e_{L56} - e_{L37}; \\
e_{L35} - e_{L67}, & e_{L67} - e_{L41}; \\
e_{L46} - e_{L71}, & e_{L71} - e_{L52}; \\
e_{L57} - e_{L12}, & e_{L12} - e_{L63}; \\
e_{L61} - e_{L23}, & e_{L23} - e_{L74}; \\
e_{L72} - e_{L34}, & e_{L34} - e_{L15}; \\
e_{L13} - e_{L45}, & e_{L45} - e_{L26}; 
\end{array} \)
\end{center}

\item $su(3)$ - \{$e_{Lpq} - e_{Lmn}$: $e_{p}e_{q} - e_{m}e_{n} = 0$, p,q,m,n = 1,...,6\};

\item $su(3)$ explicitly (this is the Lie algebra of $SU(3)$, the stability group of $e_{7}$,  
a subgroup of $G_{2}$): 
\begin{center}
\( \begin{array}{ll}
e_{L24} - e_{L56}; &  \\
e_{L35} - e_{L41}; & \\
e_{L46} - e_{L52}; &  \\
e_{L12} - e_{L63}; & \\
e_{L61} - e_{L23}; &  \\
e_{L34} - e_{L15}; & \\
e_{L13} - e_{L45}, & e_{L45} - e_{L26}.
\end{array} \)
\end{center}

\end{itemize}

The spinor space of $\textbf{T}_{L} \simeq \textbf{C} \otimes \mathcal{CL}(1,9)$ is $^{2}\textbf{T}$.  
This can be interpreted \textbf{[1]} as the direct sum of a family of leptons and quarks, and its 
antifamily (Dirac spinors, including the righthanded neutrino).  The standard symmetry, 
$U(1) \times SU(2) \times SU(3)$, was derived.  Here we shall rederive this symmetry from a 
different and more accessible direction.    The steps taken will be: \\
1. Reduce the spinor space \textbf{T} to \textbf{P} (equivalent to reducing (1,9)-Dirac spinors 
to (1,3)-Dirac spinors); \\
2. Carry this reduction to $\textbf{T}_{L}(2)$ 
and see what symmetries (bivectors) survive.  To make things easier, and because it's all that is 
necessary, we'll do this by focusing on the "Pauli" algebras, 
$\textbf{T}_{L} \simeq \textbf{C} \otimes \mathcal{CL}(0,8)$ and 
$\textbf{P}_{L} \simeq \textbf{C} \otimes \mathcal{CL}(0,2)$, 
and their respective spinor spaces, \textbf{T} and \textbf{P}.

% -----------------------------------------------------------------------------------------
\newpage

Projection operators and some of their actions:
\begin{equation}
\rho_{\pm} = \frac{1}{2}(1 \pm ie_{7}); \,\, \rho_{L\pm} = \frac{1}{2}(1 \pm ie_{L7}); 
\,\, \rho_{R\pm} = \frac{1}{2}(1 \pm ie_{R7}); 
\end{equation}
(note that $\rho_{+}\rho_{-} = 0; \,\, \rho_{+} + \rho_{-} = 1$);
\begin{equation}
\rho_{L\pm}[e_{7}] = \rho_{\pm}e_{7} = \rho_{R\pm}[e_{7}] = e_{7}\rho_{\pm} = \mp i \rho_{\pm};
\end{equation}
\begin{equation}
\rho_{\pm}e_{p}\rho_{\pm} = \rho_{L\pm}\rho_{R\pm}[e_{p}] = 0, \, p=1,...,6.
\end{equation}

In the context developed in \textbf{[1]} the spinor spaces $\rho_{+}\textbf{T}$ and 
$\rho_{-}\textbf{T}$ are the family and antifamily parts of 
$\textbf{T} = \rho_{+}\textbf{T} + \rho_{+}\textbf{T}$.  This partial reduction of \textbf{T} is 
insufficient: $\rho_{\pm}\textbf{T} \ne \textbf{P}$.  We need one further step (performed on the 
family half of \textbf{T}):
\begin{equation}
\rho_{+}\textbf{T}\rho_{+} = \rho_{L+}\rho_{R+}[\textbf{T}] = \rho_{+}\textbf{P}.
\end{equation}
The corresponding reduction on the Pauli algebra $\textbf{T}_{L}$ is:
\begin{equation}
\textbf{T}_{L} \longrightarrow  \rho_{R+}\rho_{L+}\textbf{T}_{L}\rho_{L+}\rho_{R+}.
\end{equation}

A 1-vector basis for $\mathcal{CL}(0,8)$ in $\textbf{T}_{L}$, and this same set after reduction, is:
\begin{equation}
\{ie_{L7}q_{Lr}, \,\, e_{Lp}: \,\, r=1,2, \,\, p=1,...,6\} \longrightarrow 
\rho_{L+}\rho_{R+}\{q_{Lr}: \,\, r=1,2\},
\end{equation}
a 1-vector basis for $\mathcal{CL}(0,2)$ in $\textbf{P}_{L}$.  In the broader Dirac context this 
is equivalent to reducing the 1-vectors of $\mathcal{CL}(1,9)$ to those of $\mathcal{CL}(1,3)$.

The 2-vectors of $\mathcal{CL}(1,9)$ form a basis for $so(1,9)$, the Lie algebra of the Lorentz 
group of (1,9)-spacetime.  Using the 1-vector basis in (6), a 2-vector basis for $\mathcal{CL}(0,8)$ 
(basis for $so(8)$) is:
\begin{equation}
\{q_{L3}, \,\, e_{Lpq}, \,\, ie_{Lp7}q_{Lr}: \,\, r=1,2, \,\, p,q=1,...,6\}.
\end{equation}
We'll reduce this set in steps.  First,
\begin{equation}
\rho_{L+}\{q_{L3}, \,\, e_{Lpq}, \,\, ie_{Lp7}q_{Lr}: \,\, r=1,2, \,\, p,q=1,...,6\}\rho_{L+} = 
\rho_{L+}\{q_{L3}, \,\, e_{Lpq}\},
\end{equation}
a basis for $so(2) \times so(6)$.  The next step is:
\begin{center} \(
\rho_{R+}\rho_{L+}\{q_{L3}, \,\, e_{Lpq}: \,\, r=1,2, \,\, p,q=1,...,6\}\rho_{R+} = 
\rho_{R+}\rho_{L+}\{q_{L3}, \,\, ?\}.
\) \end{center}
In particular we need to look at 
\begin{equation}
\rho_{R+}e_{Lpq}\rho_{R+}.
\end{equation}
We'll look at some examples, with $e_{p}e_{q} \ne e_{7}$, then $e_{p}e_{q} = e_{7}$,  
which will give us the general result.  Consider 
$\rho_{R+}e_{L12}\rho_{R+}$.  Because $\rho_{L+}$ commutes with $e_{Lpq}, p,q \ne 7$, 
$\rho_{L+}e_{Lpq}\rho_{L+} = \rho_{L+}e_{Lpq}$.  But $\rho_{R+}$ does not commute with these 
$e_{Lpq}$.  To see what it does we'll re-express our chosen element $e_{L12}$ as 
% -----------------------------------------------------------------------------------------
\newpage
\begin{equation}
e_{L12} = \frac{1}{2}(-e_{R12} + e_{R4} + e_{R63} + e_{R57})
\end{equation}
(see \textbf{[1]}).  $\rho_{R+}$ commutes with $e_{R12}$ and $e_{R63}$, but becomes $\rho_{R-}$ 
when commuted with 
$e_{R4}$ and $e_{R57}$ (recall: $\rho_{R+}\rho_{R-} = 0$).  Therefore, 
\begin{equation}
\rho_{R+}e_{L12}\rho_{R+} = \frac{1}{2}\rho_{R+}(-e_{R12} + e_{R63}) = 
\frac{1}{2}\rho_{R+}(e_{L12} - e_{L63})
\end{equation}
(see \textbf{[1]} for the last equality).

Finally we need to look at the three terms $e_{Lpq}$ for which $e_{p}e_{q} = e_{7}$.  These 
can be re-expressed:
\begin{equation}
\begin{array}{c}
e_{L13} = \frac{1}{2}(-e_{R13} + e_{R7} + e_{R26} + e_{R45}), \\ 
e_{L26} = \frac{1}{2}(+e_{R13} + e_{R7} - e_{R26} + e_{R45}), \\ 
e_{L45} = \frac{1}{2}(+e_{R13} + e_{R7} + e_{R26} - e_{R45}).
\end{array}
\end{equation}
All terms in (12) commute with $\rho_{R+}$, so 
\begin{equation}
\rho_{R+}\{e_{L13}, \, e_{L26}, \, e_{L45}\}\rho_{R+} = \rho_{R+}\{e_{L13}, \, e_{L26}, \, e_{L45}\}.
\end{equation}
Another basis for the space spanned by the set $\rho_{R+}\{e_{L13}, \, e_{L26}, \, e_{L45}\}$  is:
\begin{equation}
\rho_{R+}\{e_{L13} - e_{L26}, \,\, e_{L26} - e_{L45}, \,\, e_{L13} + e_{L26} + e_{L45}\}.
\end{equation}
But $e_{R7} = \frac{1}{2}(-e_{L7} + e_{L13} + e_{R26} + e_{R45})$, so 
\begin{equation}
e_{L13} + e_{L26} + e_{L45} = e_{L7} + 2e_{R7}, 
\end{equation}
which commutes with all the other surviving elements, and therefore generates a copy of $U(1)$.  

Taken all together and generalized over all $e_{Lpq}$ these results imply
\begin{equation}
\begin{array}{l}
\rho_{R+}\rho_{L+}\{q_{L3}, \,\, e_{Lpq}, \,\, ie_{Lp7}q_{Lr}: \,\, r=1,2, \,\, p,q=1,...,6\}\rho_{L+}\rho_{R+} \\
\,\,\, = \rho_{R+}\rho_{L+}\{q_{L3}, \,\, e_{Lpq}\}\rho_{R+} \\
\,\,\, = \rho_{L+}\rho_{R+}\{q_{L3}, \,\, e_{L7} + 2e_{R7}, \,\, e_{Lpq} - e_{Lmn}: \, 
p,q,m,n = 1,...,6, \,\, e_{p}e_{q} - e_{m}e_{n} = 0\}. \\
\end{array}
\end{equation}
This is a basis for (see "Some Lie algebras and their bases" above) 
\begin{equation}
so(2) \times u(1) \times su(3).
\end{equation}
The presence of the $U(1)$ generator $e_{L7} + 2e_{R7}$ ensures that the quarks charges will be 
$-\frac{1}{3}$ and $\frac{2}{3}$, given an electron charge of $-1$.  The $so(2)$ above is part of 
$so(1,3)$, a Lorentz generator.

Including the $su(2)$ from $\textbf{Q}_{R}$, the total symmetry reduction is
\begin{equation}
so(1,9) \times su(2) \longrightarrow so(1,3) \times (u(1) \times su(2) \times su(3)).
\end{equation}
As mentioned, with respect to the "internal" symmetry $u(1) \times su(2) \times su(3)$, the 
spinor space $^{2}\textbf{T}$ transforms as the direct sum of a family and antifamily of 
quarks and leptons (see \textbf{[1]} for the derivation of charges).

% -----------------------------------------------------------------------------------------
\newpage
Final notes:
\begin{itemize}

\item $u(1) \times su(2) \times su(3)$ is an internal symmetry, meaning it commutes with 
$so(1,3)$.  However, $su(3)$ is a subalgebra of $so(1,9)$.  It is a  
spacetime symmetry in this larger context; $su(2)$ is not.

\item There are two other important distinctions to be briefly mentioned.  A full resolution 
of the identity of \textbf{T} (the $\rho_{\pm}$ resolve the identity of $\textbf{C} \otimes 
\textbf{O}$) contains four members.  With these \textbf{T} can be reduced to the 
$su(2)$ vector level, but only to the $su(3)$ multiplet level.

\item And the idempotents of the resolution are $su(3)$ invariant, but not $su(2)$ invariant.

\end{itemize}

These distinctions implied in \textbf{[1]} that $su(2)$ breaks and is chiral, while 
$su(3)$ is exact and nonchiral. \\ \\

\textbf{References}: \, \\
\vspace{.25in}

[1] G.M. Dixon, \textit{Division Algebras: Octonions, Quaternions, Complex} 

\,\,\, \textit{Numbers, and the 
Algebraic Design of Physics}, (Kluwer, 1994). \\
\vspace{.25in}

[2] G.M. Dixon, www.7stones.com/Homepage/history.html

\end{document}